\documentclass[useAMS,usenatbib]{mn2e}
\usepackage{psfig,amssymb}
\usepackage{graphicx}
\usepackage{times}
\newcommand {\lsim}{\mbox{$\:\stackrel{<}{_{\sim}}\:$} }
\newcommand {\gsim}{\mbox{$\:\stackrel{>}{_{\sim}}\:$} }

\def\msun{\mbox{M$_\odot$}}
\def\Dtrans{\mbox{$D_{\rm trans}$}}
\def \ion#1#2{#1~{\sc #2}}

\def\beq{\begin{equation}}
\def\eeq{\end{equation}}
\def \Sec#1{{Sec\-tion~\ref{s:#1}}}
\def \Tab#1{{Table~\ref{t:#1}}}
\def \Eq#1{{Eq.~(\ref{e:#1})}}     % equation reference
\def \EQN#1{\label{e:#1}}        % eqn labelling a la Texsis
\def \Fig#1{{Fig.~\ref{f:#1}}}   % figure reference
\def\ga{\mathrel{\raise.3ex\hbox{$>$\kern-.75em\lower1ex\hbox{$\sim$}}}}
\def\la{\mathrel{\raise.3ex\hbox{$<$\kern-.75em\lower1ex\hbox{$\sim$}}}}

\begin{document}

%\title[Short title]{Title}
%\author[Short author list]{Author list}
%\title{Cosmological perspectives on very metal-poor stars}
\title[Population~III stars and cosmic chemical evolution]{Influence of Population~III stars on cosmic chemical evolution}
\author[Rollinde et al.]
  {E. Rollinde$^1$\thanks{E-mail: rollinde@iap.fr (ER); vangioni@iap.fr (EV); 
	dmaurin@lpnhe.in2p3.fr (DM); olive@physics.umn.edu (KAO); daigne@iap.fr (FD);
	silk@astro.ox.ac.uk (JS); vincentf@iap.fr (FHV)}, 
	 E. Vangioni$^1$\footnotemark[1], D. Maurin$^{2,1}$\footnotemark[1],
	 K.~A. Olive$^3$\footnotemark[1], F. Daigne$^1$\footnotemark[1]\thanks{Institut
	 Universitaire de France.}, J. Silk$^{1,4}$\footnotemark[1], 
	\newauthor
	and F.~H. Vincent$^1$\footnotemark[1]\\
  $^1$Institut d'Astrophysique de Paris, UMR7095 CNRS, Universit\'e Pierre et Marie Curie, 98 bis bd Arago, 75014 Paris, France\\
  $^2$Laboratoire de Physique Nucl\'eaire et Hautes Energies, CNRS-IN2P3/Universit\'es Paris VI et Paris VII, 4 place Jussieu, Tour 33, 75252 Paris Cedex 05, France\\
	$^3$William I. Fine Theoretical Physics Institute, School of Physics and Astronomy, University of Minnesota, Minneapolis, MN 55455, USA\\
	$^4$Department of Physics, Oxford University, Keble Road, Oxford OX1 3RH, UK
	}

\pagerange{\pageref{firstpage}--\pageref{lastpage}} \pubyear{Xxxx}
\date{Accepted Xxxx. Received Xxxx; in original form Xxxx}
\label{firstpage}

\maketitle

\begin{abstract}
New observations from the Hubble ultra deep field suggest that the
star formation rate at $z>7$ drops off faster than previously thought.
Using a newly determined star formation rate for the normal mode of 
Population II/I stars  (PopII/I), including this new
 constraint,  we compute the  Thomson scattering optical depth
and find a result that is marginally consistent with WMAP5 results.
We also reconsider the role of Population III stars (PopIII) in light
of cosmological and stellar evolution constraints. While this input may be needed 
for reionization, we show that it is essential in order to account for cosmic chemical evolution in the early Universe.
We  investigate the consequences of PopIII stars on the local
metallicity distribution function of the Galactic halo (from the recent
Hamburg/ESO survey of metal-poor stars) and on the evolution
of abundances with metallicity (based on the ESO large program
on very metal-poor stars), with special emphasis on carbon-enhanced
metal-poor stars.
The metallicity distribution function shape is well reproduced at
low iron abundance ([Fe/H]$\gtrsim-4$), in agreement with other studies.
However, the Hamburg/ESO survey hints at a sharp decrease of the
number of low-mass stars at very low iron abundance, which is
not reproduced in models with only PopII/I stars.
The presence of PopIII stars, of typical masses 30-40 \msun,
 helps us to reproduce this feature, leading
to a prompt initial enrichment before the onset of PopII/I stars.
The metallicity at which this cut-off occurs is  sensitive to the
lowest mass of the massive PopIII stars, which makes the metallicity
distri\-bution function a promising tool to constrain this population.
Our most important results show that the nucleosynthetic yields of PopIII stars lead
to abundance patterns in agreement with those observed
in extremely metal-poor stars. This can be demonstrated by 
the transition discri\-mi\-nant (a criterion
for low-mass star formation taking into account the cooling
due to \ion{C}{ii} and \ion{O}{i}).
In this chemical approach to cosmic evolution, PopIII stars prove
to be a compulsory ingredient, and extremely metal-poor stars are
inevitably born at high redshift.
\end{abstract}


%%%%%%%%%%%%%%%%%%%%%%%%%%%%%%%%%%%
%%%%%%%%%%%%%%%%%%%%%%%%%%%%%%%%%%%
\section{Introduction}
%%%%%%%%%%%%%%%%%%%%%%%%%%%%%%%%%%%
\label{s:introduction}

Recent data from the Hubble Ultra Deep Field (HUDF) have allowed the
determinations of the cosmic star formation rate (SFR) to be extended from redshifts $z\sim4$ up to
$z\sim7-8$ \citep{2007ApJ...670..928B}.  In the context of the  $\Lambda$CDM  model, which has by and
large  been confirmed by the observation of anisotropies in the diffuse cosmological
microwave background (CMB) by WMAP \citep{2007ApJS..170..377S,2009ApJS..180..306D}, a remaining challenge lies
in understanding the first structures, including galaxies and stars. This step requires either direct
observations (of black holes, gamma-ray bursts\ldots) or indirect constraints through
luminosity functions, metal pollution and relic metal-poor stars (e.g.
\citealt{2006ApJ...641....1T,2007MNRAS.381..647S}).  Our work develops the latter approach, and focuses on
the nucleosynthesis pollution from the first stars and their consequences for relic metal-poor
stars.

The role of massive PopIII stars as ionization sources at high redshift is poorly understood.
Indeed, their role has been clouded by the fact that the
measured value of the Thomson
scattering optical depth (from $z=0$ to the redshift
of emission of the CMB) has decreased
significantly from WMAP1 to WMAP5. Current measurements imply a redshift
for an instantaneous reionization of
$z=11.0\pm1.4$ ($\tau=0.087\pm0.017$) \citep{2009ApJS..180..306D}. In this context, the new
observations of the cosmic SFR at high redshift bring a fresh perspective on the role played by PopIII stars.  Since PopIII stars have a very specific impact on
nucleosynthesis, it  is useful to incorporate constraints from stellar
observations.
From the nucleosynthetic point of view, halo stars have long been used to
constrain galactic chemical evolution models, but were quite disconnected from cosmological
models. Yet  PopIII stars might be the primary source for the early metal enrichment of the 
interstellar medium (ISM) as well as for the intergalactic medium (IGM). The different abundance
patterns observed in extreme metal-poor stars (EMPS) may well be explained in terms of these
stars \citep{2007ApJ...663..687Y,2008ASPC..393...63F}.
It has been shown \citep{2003Natur.425..812B} that  the abundances of ionized carbon and neutral atomic
oxygen are important for the transition from PopIII to
PopII/I. \cite{2007MNRAS.380L..40F} have  defined
a  transition discriminant, \Dtrans,  and we show
 below that this quantity clearly reveals the nucleosynthetic imprint of PopIII
stars.

The paper is organized as follows. In \Sec{models}, we summarize our model for the
global chemical and cosmic evolution. New constraints related to the SFR and reionization are
described in \Sec{SFR}. 
We first reconsider the reionization constraint from WMAP5 data in terms of a
single normal mode of PopII/I star formation, by fitting the SFR to the most recent data
\citep{2007ApJ...670..928B}. We compare this result to a model which includes a contribution from
PopIII stars. 
 Stellar observations are described in \Sec{stellar}. 
%
%  We use individual stars of the ESO-LP
% 'First Stars'  \citep{2004A&A...416.1117C} and peculiar extremly-low
% metallicity stars and some of their abundances (Fe, O, C, Si) to address this
% issue.
 To address these
issues, we use some of the abundances (Fe, O, C, Si)
of individual stars of the European Southern Observatory Large Program (ESO-LP)
'First Stars'  \citep{2004A&A...416.1117C} and of peculiar extremely-low
metallicity stars.
 We also compute the metallicity distribution function (MDF, number of observed low-mass stars at a specific metallicity) and
compare this to recent observations  \citep{2008arXiv0809.1172S}.
A grid study of the model parameters
allows us to draw confidence contours for the SFR of PopIII stars and on the transition
discriminant in \Sec{grid}. 
We also consider  abundances derived from 1D and 3D model atmospheres and compare results in
each case. Our best model is used as an illustration in Sections~\ref{s:SFR} and
\ref{s:stellar}.
We summarize our results and conclude in \Sec{conclusion}.
%

%%%%%%%%%%%%%%%%%%%%%%%%%%%%%%%%%%%%%%%%
%%%%%%%%%%%%%%%%%%%%%%%%%%%%%%%%%%%%%%%%
\section{A global cosmic evolutionary model}
%%%%%%%%%%%%%%%%%%%%%%%%%%%%%%%%%%%%%%%%%
\label{s:models}

Any model for star formation in a cosmological context requires the inclusion of a model for
dark matter structure formation, accretion and outflow of baryonic matter. Numerical
simulations consistently follow the dark matter and baryonic components. Merger trees have
also been used extensively. These models have the ability to probe different IMFs and
critical metallicities. However, as yet, none of these methods follow individual element
abundances. 
In a complementary approach, we have developed a detailed model of cosmological chemical
evolution \citep{2006ApJ...647..773D}, using a simplified description of non linear structures, based on the standard
Press-Schechter (PS) formalism \citep{1974ApJ...187..425P,2001MNRAS.321..372J,1999MNRAS.308..119S}. 

One of the main
advantages over semi-analytical or fully resolved simulations (that take up a lot of computer
time) is the ability to probe a large region of parameter space
and to follow abundances of individual elements (we use only Fe, C
and O abundances in this study). Such simplified models
are successfully used to tackle specific questions related to early star formation and
reionization \citep{2008MNRAS.384.1414K,2009MNRAS.396..535H,2009arXiv0902.0853B}. These models only give average
quantities, but they can reproduce all salient features of semi-analytical approaches. They
are also supported by the lack of scatter in  chemical observations of stars, and the fact
that the MDF from Galactic halo field stars is statistically
very similar to that of
Galactic globular cluster systems and the stellar population of the nearest dSph satellites
of the Galaxy \citep{2008arXiv0809.1172S,2008ApJ...685L..43K,2009arXiv0902.2395F}.
However, note that the Galactic globular cluster MDF drops to zero at
[Fe/H]=-2.4. Presently, it is unclear where the cut-off is for the
stellar population of dSph.

We consider two distinct modes of star formation: a normal mode of PopII/I and  a massive
mode of PopIII stars. The observed SFR and element abundances at
redshift $z\lesssim 6$ (\Sec{stellar}) are
accommodated by the normal mode with a Salpeter IMF \citep{2006ApJ...647..773D}, while
the WMAP5 integrated optical depth is marginally reproduced (\Sec{SFR}).
 For PopIII stars, we
perform a grid analysis of a few key parameters to check their relevance
 for the chemical abundances of very metal-poor stars.
Throughout this paper, a primordial power spectrum with a power law index $n=1$ is assumed
and we adopt the cosmological parameters of the so-called concordance model
\citep{2007ApJS..170..377S,2009ApJS..180..306D}, i.e. $\Omega_m=0.27$, $\Omega_\Lambda=0.73$, $h=0.71$ and
$\sigma_8=0.9$.

\subsection{Description of the model}
\label{s:description}

Baryons are divided into three reservoirs. Two reservoirs account for the matter in collapsed structures: the gas (hereafter ``ISM'')
and the stars and their remnants (hereafter ``stars''). The third
reservoir represents baryons in the medium between these structures
(hereafter ``IGM''). The evolution of the mass of each reservoir is
governed by equations (1) and (2) in Daigne et al. 2006 (see also figure
1 in the same paper), which depend on different fluxes: (i) baryon
accretion from the IGM to the ISM, due to the structure formation
process. This term is computed at each redshift in the framework of the
Press \& Schechter formalism (see equation (6) in Daigne et al. 2006),
assuming that collapsed structures correspond to dark matter halos with
a minimum mass $M_\mathrm{min}$ (which, for simplicity, is kept constant
over the entire evolution); (ii) baryon ejection from the ISM into the
IGM, associated with outflows from the collapsed structures. Such
outflows are powered by a fraction $\epsilon$ of the kinetic energy
released by SN explosions. The corresponding mass flux is evaluated by
assuming that the typical velocity of these outflows  scales as the mean
escape velocity at redshift $z$. This velocity is computed by averaging
the escape velocity over the distribution of the mass of dark matter
halos above $M_\mathrm{min}$ at redshift $z$. Outflows become less
efficient with time as the typical mass of collapsed structures
increases; (iii) the third flux is associated with the star formation
process and transfers baryons from the ISM to stars. It can easily be
computed once a star formation rate $\Psi(z)$ has been specified. Our
assumptions regarding this SFR are detailed below; (iv) the final flux
stands for mass ejection from stars to the ISM, associated with stellar
winds and explosions. It is computed from the SFR and the IMF. We do not
assume instantaneous recycling and therefore take into account the
lifetimes of stars. The stellar data we use are described in the next
subsection and are dependent on both the initial mass and metallicity of the stars.

In addition to the evolution of the mass, we also compute the chemical evolution in the ``ISM'' and the ``IGM''. For any chemical element $i$, the mass fractions in the ISM and the IGM, $X_{i}^\mathrm{ISM}$ and $X_{i}^\mathrm{IGM}$,  are followed as a function of the redshift. The corresponding differential equations can be found in Daigne et al. 2004 (equations (6) and (7)) and are derived from the previous mass fluxes assuming that (i) the chemical composition of the accretion flow related to the structure formation process is the composition of the IGM at redshift $z$; (ii) the chemical composition of the outflows escaping collapsed structures is the composition of the ISM at redshift $z$; (iii) this composition is also the composition of stars that form at the same redshift; (iv) finally the composition of the matter ejected by stars into the ISM is given by stellar yields. These yields are detailed in the next subsection. Once the chemical evolution is known, it can be inverted to estimate the redshift of formation of a star from its observed abundance, as explained in \Sec{Ztoz}.

The models considered here are bimodal. Each model contains a normal mode with stellar masses
between 0.1 M$_{\odot}$ and 100 M$_{\odot}$, with an IMF with a near Salpeter slope. The SFR
of the normal mode peaks at $z \approx 3$.  In addition, we allow for a massive component,
which dominates star formation at high redshift. We primarily focus on stars with 40-100
$M_\odot$, which terminate as type II supernovae. 
%By varying this mass range, we can also
%inspect the role played by this massive mode.  
The nucleosynthetic pattern of
pair-instability SN (PISN) is briefly commented on in the conclusions.

In addition to the SFR, the IMF and stellar data, this simple
model introduces only two other parameters: the minimum mass of the
dark matter halo in collapsed structures, $M_\mathrm{min}$, and the
efficiency of supernovae to power outflows from collapsed structures,
$\epsilon$. The impact of these two parameters has been discussed in
detail in Daigne et al. 2006. In the present paper, we adopt a typical
scenario with $M_\mathrm{min}=10^{7}\ \mathrm{M_\odot}$  and
$\epsilon=3\times 10^{-3}$. This minimum mass,  $M_\mathrm{min}$,
is also 
 deduced from hydrodynamical simulations at $z\sim 15$. Note that
 \cite{2009arXiv0902.3263J} consider $M_{\rm min}=10^{8} M_\odot$ at $z=$
 12.5 \citep[see
 also][]{2006ApJ...647..773D,2009arXiv0902.0853B,2009ApJ...694..879T}.
$\epsilon$ primarly impacts on the enrichment of IGM, which is not
considered in this paper.

\subsection{Yields and lifetimes}
\label{s:yields}

 The lifetimes of intermediate mass stars ($0.9<M/\msun <8$) are taken from \citet{1989A&A...210..155M} and  from
\citet{2002A&A...382...28S} for more massive stars. 
Old halo stars with masses below $\sim 0.9\, \msun$ have a lifetime  long enough to be
observed  today. They inherit the abundances of the ISM at the time of their formation.
Thus, their observed abundances reflect, 
in a complex way (due to exchanges with the IGM),
the yields of all massive stars that have exploded earlier.

The yields of stars depend on their mass and their metallicity, but not
on their status (i.e. PopII/I or PopIII). Some PopII/I stars are
massive, although in only a very small proportion since we use a steep Salpeter
IMF. PopIII stars are all very massive stars. We use the tables of
yields (and remnant types)  given in~\citet{1997A&AS..123..305V} for intermediate mass stars ($<8 \msun$),
and the tables in~\citet{1995ApJS..101..181W} for massive
stars ($8<M/\msun <40$ ). An interpolation is made between
different metallicities (Z=0 and Z=$10^{-4,\, -3,\, -2,\, -1,\, 0}$Z$_\odot$)
and we extrapolate the tabulated values beyond 40 \msun.

 \citet{2008arXiv0803.3161H} have provided new
  stellar yields in the  mass range  $10-100$ \msun, but only at zero-metallicity. 
 At zero-metallicity, and in the $40-100$ \msun mass range, we 
have checked that our extrapolated values are consistent with those
  yields. Indeed, their favored model corresponds to  very little
  mixing and a high oxygen to iron ratio, in agreement with the 
\citet{1995ApJS..101..181W} model we chose.
 We  explicitly checked that our results were not modified using either
 ~\citet{1995ApJS..101..181W} or \citet{2008arXiv0803.3161H}
at zero-metallicity.

\section{Cosmological constraints}
\label{s:SFR}

Taking into account the results of recent observations on the PopII/I SFR~\citep{2007ApJ...670..928B}, we
reanalyse the reionization capability of these stars. The WMAP5 upper limit on the Thomson
optical depth is then used to set constraints on the PopIII SFR.

\subsection{PopII/I SFR}

%SFR
\begin{figure}
\includegraphics[width=\linewidth]{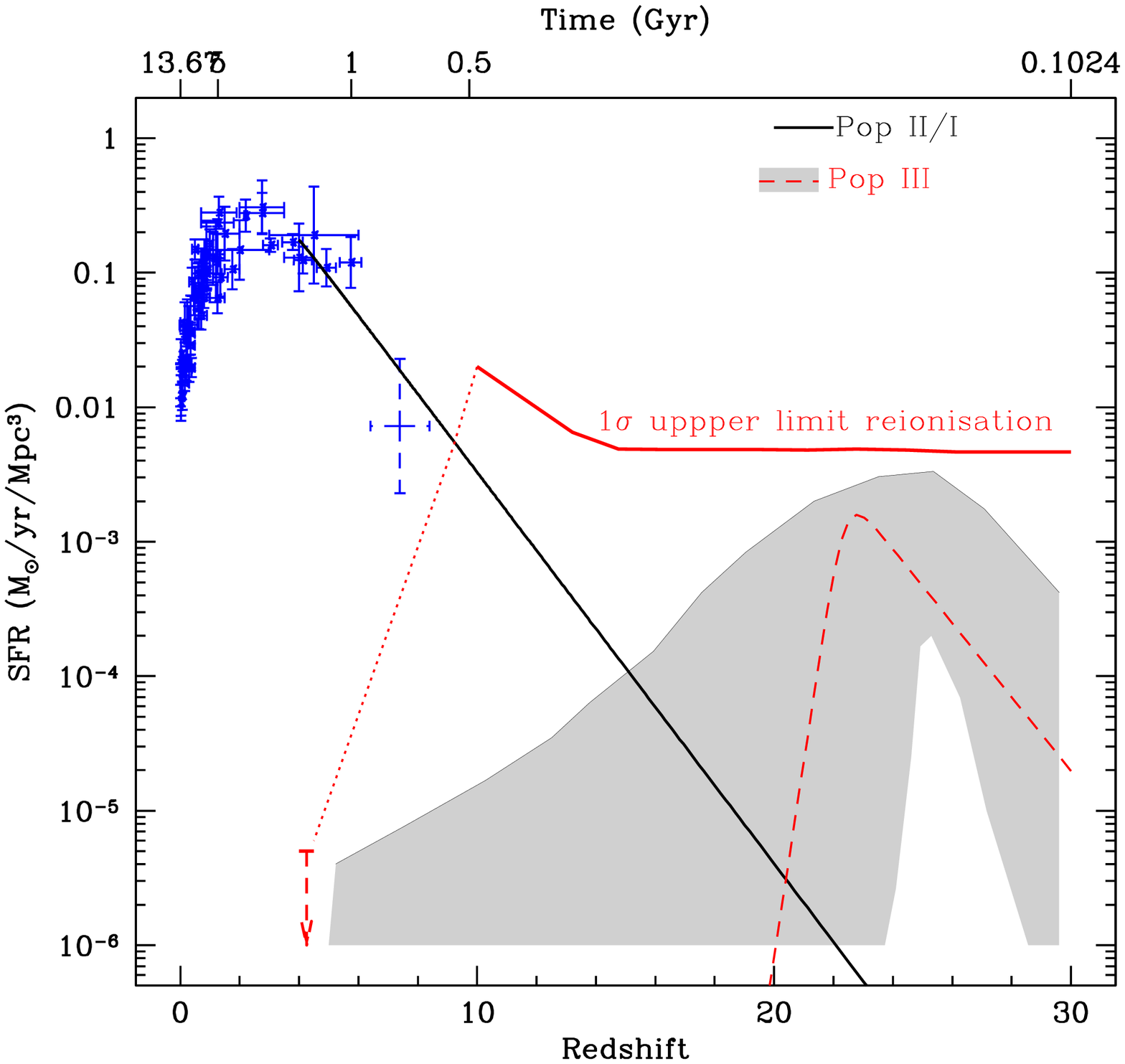}
\caption{Cosmic SFR as a function of redshift. {\em PopII/I}: The observed SFR up to $z\sim 5$ (solid blue
points) is taken from \citet{2006ApJ...651..142H}. The dashed blue data point comes from
\citet{2007ApJ...670..928B}.
 The solid black line is the extrapolated PopII/I SFR, using
 the shape given by \Eq{SFRenv}, in order to match the data at
 $z\lsim 5$ and the upper limit at $z\sim7$. 
{\em PopIII}: The recent non-observation of Ly$\alpha$-HeII
 dual emitters \citep{2008ApJ...680..100N} at
$z\sim 4-4.5$ is shown with the red arrow.
The red line corresponds to the PopIII SFR upper limit, adjusted to satisfy  the 1$\sigma$ WMAP5 upper limit on
the Thomson optical depth (see \Fig{tau}).
This  limit has to drop (dotted line)
to be reconciled with the Ly$\alpha$-HeII
 observation.  
The envelope (shaded area) contains all  models, that reproduce the
 \Dtrans\ evolution at a 95\%~CL (see \Sec{grid} and \Fig{Dtrans})
For illustration, the best model for the  PopIII SFR (including nucleosynthetic
 and cosmological constraints, described in \Sec{stellar}) is shown by the dashed red line.
This SFR is parametrized  with \Eq{SFRenv}, with the following parameters:
$\nu_{\rm III}=0.0016$  M$_\odot$ yr$^{-1}$ Mpc$^{-3}$ (astration
rate for PopIII stars), $z_{\rm m\, III}=22.8$, $a_{\rm III}=4$ and $b_{\rm III}=3.3$.}
\label{f:SFRenv}
\label{f:SFR}
\end{figure}

We fit the SFR history of PopII/I stars to the data compiled in \citet{2006ApJ...651..142H} (from $z=0$ to
5), and to the recent measurement at $z\sim 7$ by \citet{2007ApJ...670..928B} (see Fig.~\ref{f:SFRenv}).
The behaviour of the SFR at high redshift is still disputed. Some indirect measurements hint
against a decline of the SFR beyond $z\sim3$ \citep{2008ApJ...682L...9F}, but direct observations of
$z_{850}$-dropout galaxies seem to confirm a sharp decrease of the SFR at $z\sim 6-7$
\citep{2009ApJ...690.1350O}. The low level of the \citet{2007ApJ...670..928B} data  thus places strong
constraints on the PopII/I SFR.

The fit for the SFR, $\psi(z)$, is based on the mathematical form proposed in \citet{2003MNRAS.339..312S}:
\begin{equation}
\psi(z) = \nu\frac{a\exp(b\,(z-z_m))}{a-b+b\exp(a\,(z-z_m))}\,.
\EQN{SFRenv}
\end{equation}
The amplitude (astration rate) and the redshift
 %position
 of the SFR maximum are given by $\nu$ and  $z_m$ respectively, while
$b$ and $b-a$ are related to its slope at low and high redshifts
respectively. In the following, we use the subscripts II/I and
III for parameters related to PopII/I and PopIII SFRs respectively. The thick black
curve in \Fig{SFRenv} is computed with $\nu_{\rm II/I}=0.3$ M$_\odot$ yr$^{-1}$ Mpc$^{-3}$, 
$z_{\rm m\, II/I}=2.6$, $a_{\rm II/I}=1.9$ and $b_{\rm II/I}=1.2$. The parameter values chosen here differ from that in
\citet{2006MNRAS.373..128G} so as to better account for the high redshift data point \citep{2007ApJ...670..928B}.
To be conservative with respect to the reionization constraint (see below), we choose a SFR
consistent with the $1\sigma$ upper limit of this $z=7$ observation. A steeper slope only
decreases the number of ionizing photons.

\subsection{Reionization from PopII/I stars}
Having set the SFR in our model, we now compute the electron scattering optical depth for our
choice of IMF. The evolution of the volume filling fraction of ionized regions is given by:
\begin{eqnarray}
\frac{\mbox{d}Q_{{\rm ion}}(z)}{\mbox{d}z}&=&\frac{1}{n_{\rm
 b}}\frac{\mbox{d}n_{{\rm ion}}(z)}{\mbox{d}z}-\alpha_{{\rm B}}n_{{\rm b}}C(z)\nonumber\\
&&\times Q_{{\rm ion}}^{2}(z)\left(1+z\right)^{3}\left|\frac{\mbox{d}t}{\mbox{d}z}\right|\mbox{\ ,}
\end{eqnarray}
where $n_{\rm b}$ is the comoving density in baryons, $n_{{\rm ion}}(z)$ the comoving density
of ionizing photons,  $\alpha_{{\rm B}}$ the recombination coefficient,  and $C(z)$ the
clumping factor. This factor is  taken from \citet{2006MNRAS.373..128G} and  varies from a value of 2 at $z\leq20$ to a constant value of 10 for 
$z<6$. The
escape fraction,  $f_{\rm esc}$, is set to 0.2 for both PopIII and PopII/I SN, while
\citet{2006MNRAS.373..128G} take two different values. The number of ionizing photons  for massive
stars is calculated using the tables given in \citet{2002A&A...382...28S}. Finally, the Thomson optical
depth is computed as in \citet{2006MNRAS.373..128G}:
\begin{equation}
\tau =c\sigma_{{\rm T}}n_{\rm b} \int_{0}^{z}dz'\,Q_{{\rm ion}}(z')\left(1+z'\right)^{3}\left|\frac{\mbox{d}t}{\mbox{d}z'}\right|\mbox{\ ,}
\end{equation}
where $z$ is the redshift of emission, and $\sigma_{{\rm T}}$ the Thomson scattering
 cross-section.

\begin{figure}
\includegraphics[width=\linewidth]{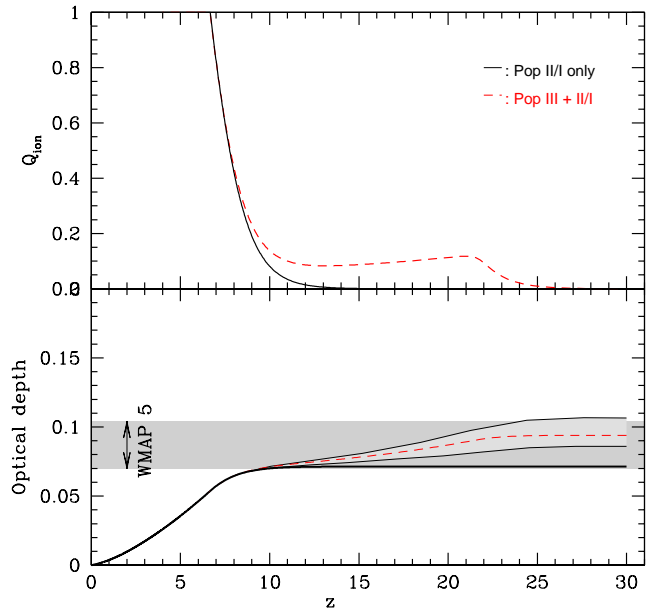}
\caption{Histories of volume filling fraction of ionized regions, Q$_{\rm ion}$ (top panel)
and the implied optical depth (lower panel) with and without PopIII stars (dashed red and
solid black respectively).
 The envelope in the lower panel (light shaded area)  shows the maximum and
 minium values of optical depth reached by  the different models that
 reproduce the chemical abundance constraints at a 95\%~CL (see \Sec{grid}).
 The lower panel also shows the range of allowed values for the
optical depth as measured by WMAP5 \citep{2009ApJS..180..306D}.}
\label{f:tau}
\label{f:tauenv}  
\end{figure}

In the upper panel of \Fig{tau}, we show the predicted evolution of the volume filling factor
of ionized regions, $Q_{\rm ion}$  as a function of redshift. In the lower panel, we show the
integrated optical depth from $z=0$ to $z$. These results can be compared to Fig. 5 of
\citet{2006MNRAS.373..128G}. We see that the normal mode (black solid curve) is contained in the
$1\sigma$ WMAP limit (grey band) for the observed value for the Thomson optical depth.

\subsection{Reionization from PopIII stars}
\label{s:tauIII}
A massive PopIII mode at high redshift makes it easier to reproduce the optical depth at high
redshift. For illustrative purposes, the history of reionization is shown in \Fig{tau} for a
typical SFR (red dashed line in \Fig{SFRenv}) found to best meet 
the local constraints (discussed in
\Sec{grid}). The parameters of the best fit model for PopIII SFR are given in the caption of \Fig{SFRenv}.
 The associated history of reionization  is consistent
with a two-step model of reionization as invoked in \citet{2009ApJS..180..306D}, who also comment that
{\em the WMAP5 data suggests a more gradual process with reionization beginning perhaps as
early as ~20.} In particular, the PopIII SFR allows $Q_{\rm ion} \ne 0$ at $z>10$.
The bimodal SFR is consistent with recent general scenarios that  have emerged from
hydrodynamic simulations and analytical calculations  (e.g. \citealt{2004ARA&A..42...79B,2008AIPC..990..405G}). These
studies have shown that, due to the re-collapse of the structures after disruption by
massive SNe,  low-mass star formation may have been delayed by more than a Hubble time or up
to a redshift $\sim$ 10. This can be understood as a statistical description of the 
mechanical feedback implemented in merger trees \citep{2007MNRAS.381..647S} where a time delay
between PopIII and PopII/I  appears at each individual branch. 

Using a grid analysis on the parameters of \Eq{SFRenv} for PopIII stars 
(see details in \Sec{grid}), we are able to derive an
upper limit on the SFR. There is no lower limit as PopII/I stars alone are consistent with the
WMAP reionization lower limit. The 68\%~CL upper limit on the SFR is reported as a
solid red line in \Fig{SFRenv} that we stop below $z<10$. At lower redshift, the most stringent
constraint is set by the recent non-observation of Ly$\alpha$-HeII dual emitters
\citep{2008ApJ...680..100N},  which gives an upper limit for
the PopIII SFR of $5\cdot10^{-6}$~M$_\odot$~yr$^{-1}$ Mpc$^{-3}$ at
$z\sim 4-4.5$ (red arrow). The dotted line reconciles this
observational constraint with the WMAP5 upper limit.
 Note
that within the region allowed by the cosmological constraint (bottom right), the shape of
the SFR of PopIII remains completely unspecified (based on the reionisation constraint), as
the relevant quantity is related to the integral over $z$ of the SFR. We show in the following
section that local observations require the existence of PopIII stars, and set
stronger constraints on the SFR and its lower limit.

\section{Local observations}
\label{s:stellar}

The average mass of the structures in the PS formalism is close
to the total mass of the Milky Way ($10^{12}$ M$_\odot$) at $z=3$, which
corresponds to the epoch of the peak of the global SFR. Our average description of the cosmic history is thus
expected to be representative of the evolution of the Galaxy. In this section, local
observations are analyzed in a global cosmological context.

\subsection{Metallicity Distribution Function (MDF)}
\label{s:MDF}

%MDF
\begin{figure}
\includegraphics[width=\linewidth]{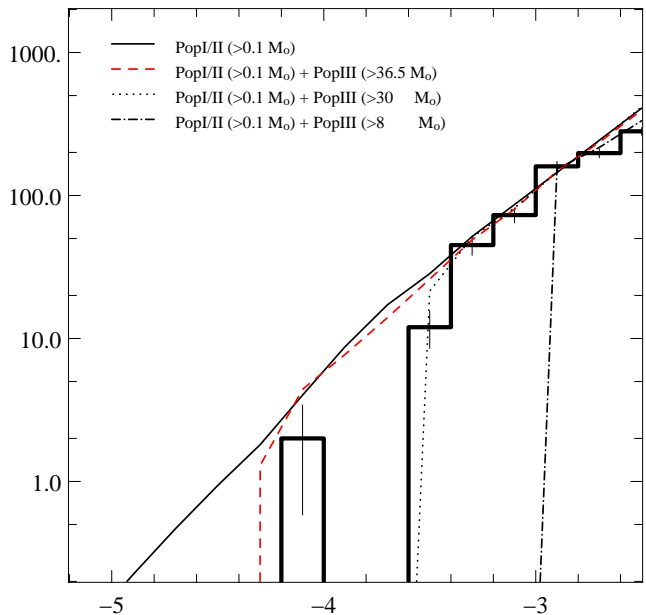}
\caption{Comparison of the observed MDF (\citealt{2008arXiv0809.1172S}, thick histogram) for the
galactic halo stars with model calculations of the normal mode of star  formation (black
solid line). Other lines show the effect of the  inclusion of PopIII stars with different
mass ranges. All distributions are normalized to match the observations at [Fe/H]=-3.}
\label{f:MDF} 
\end{figure}

The MDF is the distribution of stars at a given iron abundance, derived from observations that have
been corrected for various technical and observationnal biases. It has been
recently published by the Hamburg-ESO survey \citep{2008arXiv0809.1172S} using  the \citet{2008A&A...484..721C} catalog. The
thick line histogram in \Fig{MDF} is taken from \citet{2008arXiv0809.1172S}. As noted there, in order
to compare observed and predicted MDF, one has to take into account the modification of the
shape of the MDF by the selection of metal-poor candidates. Consequently,
a comparison to our prediction is valid for [Fe/H]$<-2.5$ only\footnote{For a given element
X, the abundance relative to the solar value is defined by
[X/H]~=~$\log_{10}($X/H$)~-~\log_{10}($(X/H)$_\odot)$.}.

 In our analysis, for a given model (SFR and IMF of PopII/I and PopIII), we count the
 number of low mass stars created at each
redshift. As the iron abundance in the ISM is also calculated as a
 function of the redshift, the metallicity of  the stellar population at each redshift is known (\Sec{description}). 
The value of the MDF in a given iron
 abundance bin,  is given  by the number  of   stars that are still shining today
 and  were created  at times when the iron abundance was within this
bin.
The MDF for several 
possible PopIII star formation histories are plotted along with the observation in \Fig{MDF}.
The solid black line corresponds to the normal mode, i.e. a
 Salpeter IMF starting at  0.1
M$_\odot$. 
 A massive mode is then added with a
 Salpeter IMF with different lower masses: 36.5 M$_\odot$ (red
dashed), 30 M$_\odot$ (black dotted) and 8 M$_\odot$ (black
dot-dashed). The minimum mass of 36.5 corresponds to our best
 model. Note that the value of the upper limit of the IMFs is not
 a key parameter, in constrast to the lower one, due to the steep slope of
 the IMF.

For $-4<$[Fe/H]$<-2.5$, the slope of the MDF is well reproduced
by PopII alone. 
The reality of the drop observed at [Fe/H]$\sim-3.6$ (compare the thin solid
black line to the thick black histogram) is still debated
(e.g., \citealt{2009arXiv0901.0617K}). The prediction of several models for the MDF is
discussed in \citet{2008arXiv0809.1172S}, where the authors conclude that none of them are able to
explain the tail at [Fe/H]$<-4$, except a stochastic enrichment model
(e.g. \citealt{2006ApJ...641L..41K}).
In our model,  as in Karlsson's model, the presence of a massive mode
implies the early production of iron, which has an impact on the low metallicity MDF: very few low
mass stars at zero or quasi-null metallicity are expected in such a scenario, i.e., 
effectively as a prompt initial enrichment (PIE). The cut-off on the MDF at low metallicity
depends on the lower mass of the massive mode. 

Given the paucity of stars with 
[Fe/H]$\lesssim -3.5$, it is difficult to accurately discriminate between different star
formation histories presently. Indeed, PopIII stars
do not modify  the
MDF at $-3.5<$[Fe/H]$< -2.5$, as long as their minimal mass is larger
than about 30 \msun.

\subsection{Global nucleosynthetic evolution}
\label{s:Dtrans}

We now focus on the giants of the ESO-LP \citep{2004A&A...416.1117C} which contain the abundances of 17
elements from C to Zn for all of the observed stars. Following the nomenclature proposed in
\citet{2005ARA&A..43..531B}, very metal-poor (VMP) stars  correspond to a metallicity
${\rm [Fe/H]}<-2$, extremely metal-poor (EMP) stars 
to  ${\rm [Fe/H]}<-3$, ultra metal-poor (UMP) stars to [Fe/H]$<-4$ and
hyper metal-poor (HMP)
stars  to ${\rm [Fe/H]}<-5$. Some UMP stars are also carbon-enhanced
metal-poor (CEMP) stars
 with [C/Fe]$>+1$.

\subsubsection{ESO-LP  and CEMP data}
The ESO-LP contains 35 VMP stars, which includes 22 EMP stars. The star
in their sample with the
next to  lowest iron abundance
(CS 22949-037; [Fe/H]~=~-3.97) is a CEMP star. Note that CD-38$^\circ$245
\citep{1984ApJ...285..622B} has [Fe/H]=-4.19; unfortunately,
\citet{2004A&A...416.1117C} and \citet{1984ApJ...285..622B} do not
provide O abundances, and only an upper limit to the abundance of C.  
One result of this survey has been the reduction of the scatter in different element
abundances as correlated against [Fe/H] (e.g Mg, Ca, Cr and Ni). According to \citet{2005ARA&A..43..531B},
this is contrary to the long-standing hypothesis that, at such  low metallicity, one observes
the nucleosynthetic products of only a few or even a single SN II
\citep{1998ApJ...507L.135S,2000ApJ...531L..33T}. The lack of scatter in these abundances could be
explained by a well-mixed ISM.

\begin{table}
\caption{{\bf Abundances of CEMP stars}}
\label{t:CEMP}
\begin{tabular}{cccc}
Name & [Fe/H] & [C/H]  & [O/H] \\
     & 1D/3D$^\star$ & 1D/3D & 1D/3D \\
\hline
HE 1300+0157 $^a$&  $-3.88$ / $-3.88$ & $-2.50$ / $-3.10$ & $-2.12$ / $-2.72$\\ 
HE 1327-2326 $^b$&  $-5.46$ / $-5.96$ & $-1.55$ / $-2.18$ & $-1.76$ / $-2.54$ \\
HE 0107-5240 $^c$&  $-5.30$ / $-5.50$ & $-1.30$ / $-2.30$ & $-3.00$ / $-3.60$\\
G-77-61 $^d$&  $-4.00$ / $-4.20$ & $-1.40$ / $-2.40$ & $-2.20$ / $-2.80$ \\
\hline
\end{tabular}

\medskip
Abundances derived from 1D and 3D model
atmospheres
$^\star$ Abundances derived from 3D/NLTE model atmospheres are used for the main  analysis (\Sec{grid}). 
 Abundances derived from 1D model atmospheres are considered in \Sec{1d3d}.  \\
$^a$ assumed to be a subgiant as in \citet{2007ApJ...658..534F}.\\ 
$^b$  \citet{2008ApJ...684..588F}.\\
$^c$ \citet{2004ApJ...612L..61B,2004ApJ...603..708C}; abundances derived
 from 3D model atmospheres are
 estimated from \citet{2006ApJ...644L.121C}.\\
$^d$ \citet{2007AJ....133.1193B,2006ApJ...644L.121C}. 
\end{table}

The CEMP stars references are retrieved from the SAGA
tool\footnote{http://saga.sci.hokudai.ac.jp/Retrieval/db.cgi} presented
in \citet{2008PASJ...60.1159S}. Abundances derived from 1D and 3D model
atmospheres and references are
gathered in \Tab{CEMP}. At
the lowest metallicity, HE~0107-5240 and
HE~1327-2326 are both HMP and CEMP stars. The very specific chemical
pattern of these stars is assumed to be related to the first stages of
star formation. 
The UMP star HE~1300+0157 with [Fe/H]=-3.88  has an abundance
pattern close to the single UMP star in the ESO-LP.
For  G77-61,  the criterion of CEMP-no star is not strictly met
(only an upper limit exists on [Ba/Fe]).
Moreover, in contrast to the other stars considered, the
abundances of G77-61 are based on medium resolution optical
([Fe/H], [C/Fe]) and near-infrared ([O/Fe]) spectroscopy. This method
can bring systematic uncertainties and possibly an over-estimate of C
 \citep{2007AJ....133.1193B}. Consequently, this star is not used in the
 analysis, but is displayed in the figures for illustrative purposes.
 There is one additional star with $-5 <$
[Fe/H] $< -4$, HE~0557-4840 \citep{2007ApJ...670..774N}. While it also has an enhanced C to Fe ratio, there is only an upper
limit on [O/H] making it difficult to place this star directly into our
analysis. 

For our main analysis, abundances derived from 3D model atmospheres are considered first. We
also discuss the impact of using 1D versus 3D model atmospheres in \Sec{1d3d}.

\subsubsection{Transition discriminant}

The transition from one star formation mode to another is often described in terms of a
critical metallicity. For example, \citet{2003Natur.425..812B} have shown the importance of
ionized carbon and neutral atomic oxygen in this transition and suggest that when
sufficiently abundant, these elements act as a trigger to lower mass star formation. 
%
%Fig.8 : Dtrans.pepsnd abundance.eps
\begin{figure}
\includegraphics[width=\linewidth]{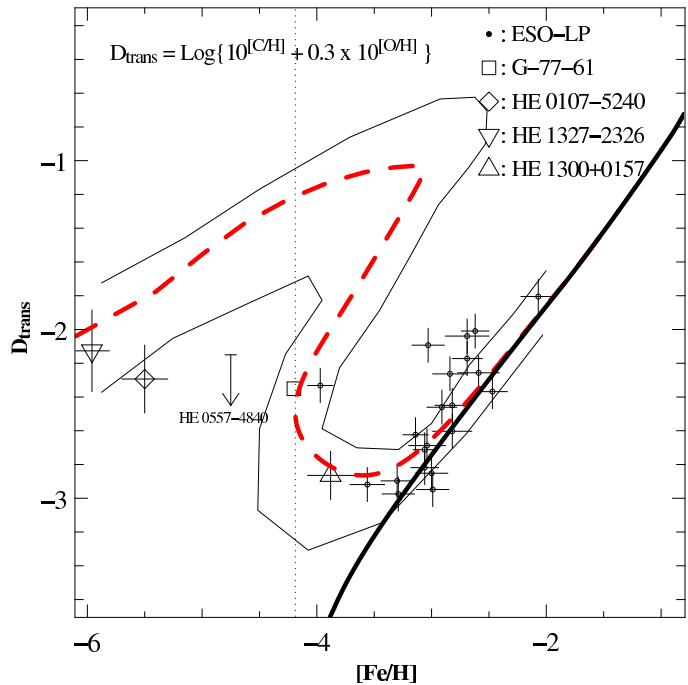}
\caption{The transition discriminant, $\Dtrans$, for metal-poor stars as a function of
[Fe/H].  The ESO-LP stellar data are shown as circles while
peculiar CEMP stars are indicated with different symbols. 
The thick black solid line corresponds to a model with PopII/I
 stars alone. The addition of PopIII
stars (35-100 \msun, our best model) 
leads to the red dashed line. The dotted vertical line
indicates the prompt initial enrichment (PIE) produced by PopIII stars
 (see \Fig{MDF}).
The envelope (thin black curves)
contains all  evolution curves predicted by models, that match
the abundance at a 95\%~CL, as determined by our $\chi^2$
analysis (see \Sec{grid}). Only stars with error bars are included
in this analysis.}
\label{f:Dtrans}
\label{f:Dtransenv} 
\end{figure}
%
%On one hand, it has often been argued that low mass stars could not form in a metal-free
%environment. On the other hand, it was recognized that the production of carbon and oxygen by
%early generations of  massive  stars would trigger the formation of low mass  stars. Since
%the iron abundance  is not the best tracer of star formation,
\citet{2007MNRAS.380L..40F}  defined the 
{\em transition discriminant}: % to take into account the predominant role of carbon and oxygen:
\begin{equation}
\Dtrans\equiv \log_{10}(10^{\rm [C/H]}+0.3\times10^{\rm [O/H]})\, .
\end{equation}
It is  argued that low-mass star formation requires $\Dtrans\gsim -3.5\pm 0.2$ 
\citep{2006ApJ...643...26S,2007MNRAS.380L..40F}.
A scatter plot of \Dtrans\ as a function of [Fe/H] is shown in \Fig{Dtrans}. All ESO-LP very
metal-poor stars (circles) and CEMP stars meet the condition $D_\mathrm{trans}\gsim -3.5$.
%The distinction between the four peculiar stars and the bulk of the ESO-LP stars is made
%clear with this representation.

\subsubsection{PopIII stars and the enrichment of CEMP stars}

Our model follows  C, O and Fe abundance evolution with $z$, and hence enables us to 
calculate  \Dtrans\ and [Fe/H] at each redshift (see
\Sec{models} for details and \citealt{2006ApJ...647..773D}). In the \Dtrans\ versus [Fe/H] scatter
plot of \Fig{Dtrans}, the evolution from $z=30$ to $z=0$ corresponds to a continuous path
starting at the lowest metallicity (left-hand side) and ending up at the highest metallicity
(right-hand side). 

A standard IMF used for the normal mode leads to the thick black-solid line. It reproduces well the
ESO-LP observations, but it is clear that  the CEMP
stars abundances are not explained in this standard scenario.
 The evolution of \Dtrans\  in our best fit model (see next section)
 which includes a massive mode is  shown as the red-dashed line. 
The SFR corresponding to this model is shown in \Fig{SFRenv}
(red-dashed line) and corresponds to the parameters given above in \Sec{tauIII}.

PopIII SN eject more C and O than PopII/I yielding
higher values of \Dtrans. This allows one to explain the abundances of the peculiar stars. As
a result, these stars must be associated with star formation occurring within
regions having a specific nucleosynthetic history at early times as suggested in
\citet{2008ASPC..393...63F}. At intermediate redshift,
 we see a decrease of both \Dtrans\ and [Fe/H].  This
corresponds to the epoch where the SFR is reduced  (see \Fig{SFRenv}). The accretion of pristine IGM
material continues, leading to the  dilution of ISM abundances. The abundance dilution allows
one to reproduce simultaneously  the observations of CEMP stars at high redshift and of ESO-LP stars at
low redshift. As PopII/I stars turn on, we reconnect to the black solid line. Finally, the
vertical dotted line indicates the PIE produced by PopIII stars as discussed already for the
MDF drop in \Sec{MDF}.
%This implies that the massive mode is specifically
%present at high redshift. Its amplitude has no effect
%on the calculated abundance ratios. It also prevents
%a large number of PopIII stars at low redshift. 

\section{Constraints on PopIII stars from local observations}
\label{s:grid}

As shown above,  PopII/I stars barely account for
the reionization of the Universe. PopIII stars are  not essential in this context,
unless the SFR of PopII/I stars at $z>7$ drops off faster than in our conservative choice (see
\Fig{SFRenv}), or the value of the escape fraction used in our study
(20\%) is overestimated. Conversely, if we turn to the local
observations and interpret them in a cosmological context, PopIII stars are mandatory from
chemical observations, especially based on the consideration of  CEMP stars.
Hence, if we consider that the abundances of CEMP stars are representative of
the nucleosynthesis of the first stars, we can perform a
statistical analysis to constrain the PopIII parameters. 
We have shown above that the few stars observed at very low metallicities 
are not sufficient to use the MDF as a constraint on PopIII models.
To reproduce the slope of the MDF at [Fe/H]$>-3.5$, we consider only minimal
 masses of 30-40 \msun\ for PopIII stars.

We first describe the method to evaluate the quality of the model.
In this process, a redshift of formation can be set to each observed
star. We then discuss the constraints set by abundances derived from 3D
model atmospheres and the uncertainties
related to 1D or 3D model atmospheres.

\subsection{Methodology}

The analysis is done in the  
\Dtrans-[Fe/H] plane, 
% (and also combined with the reionization constraint),
 using a grid approach to cover the parameter space of the model. The
parameters that we vary are $\nu_{\rm III}$ (from $2 \times 10^{-5}$ to
 $2 \times10^{-2}$ M$_\odot$ yr$^{-1}$ Mpc$^{-3}$), $z_{m {\rm III}}$ (from
 10 to 26), $a_{\rm III}$ (from 0.5 to 5), $b_{\rm III}$ (from 0.03 to 4.9) 
 and the minimum mass of the massive mode, $M_{\rm III}$ (from 30 to 39 
 M$_\odot$).

\subsubsection{Minimization procedure}

For each model, we calculate the \Dtrans -[Fe/H] curve, which
 is multi-valued for models including  PopIII stars.  
%An estimator is necessary to evaluate the quality of the match
%to the data. For each star, we calculate the  minimal distance 
%to the curve (taking into account observationnal uncertainties) .
% Our $\chi^2$ estimator is the sum of the squared distances for all stars.
%
The best fit model corresponds to the set of parameters that minimizes
%this estimator. 
a $\chi^2$ on the predicted and observed stellar iron
abundance and transition discriminant \Dtrans\ (via stellar carbon and
oxygen abundances) of a sample of 21 ESO-LP and 3 peculiar stars. 
We consider all parameters within a 95\%~confidence level (CL)
based on this $\chi^2$ to be acceptable. Therefore, in all figures, we have displayed 
 95\%~CL envelopes that contain the predicted curves of all
those acceptable models.

\subsubsection{Transforming stellar abundances to  redshift }
\label{s:Ztoz}

%Different star formation histories lead to different
%chemical evolutions for element abundances. 
In many models, the iron abundance is monotonically
increasing with time, giving a unique correspondance between [Fe/H]
and redshift. This is the classical view of 
cosmological chemical evolution, where the oldest stars are the
most metal-poor. However, in the first metal-free structures,
the nucleosynthesis of PopIII stars provide features in the evolution of [Fe/H], 
so that the previous picture does not necessarily
hold. In that case, not only the abundances of Fe, but also 
of elements such as C and O are  required in order to be able to 
assign the VMP stars to a particular redshift.

Each position along the predicted curve \Dtrans/[Fe/H] is
associated with a 
redshift. Following the procedure described above, we search for the position along this curve that
minimizes the distance to the observed values for a given star. This position gives the redshift of formation of this star. The predicted curve depends on the model considered, and therefore different models will associate 
different redshifts of formation with the same star. As we repeat this procedure for models that
satisfy the 95\%~CL on the whole sample, we can infer the allowed range of redshift
(from the minimum to the maximum redshift) for one specific star.

\subsection{abundances derived from 3D model atmospheres}

%\subsubsection{\Dtrans\ envelopes}

Models that satisfy a 95\%~CL limit are  included within the envelope
shown in \Fig{Dtransenv} (thin black lines). %The best model (used in the
%previous sections) is shown with a dashed red line. It corresponds to 
%$\nu_{\rm III}=0.00036$\, M$_\odot$ yr$^{-1}$ Mpc$^{-3}$; $z_{m {\rm III}}=24.5$; 
%$a_{\rm III}=0.5$; $b_{\rm III}=0.4$ and $M_{\rm III}=36.5$\, \msun.
%
All stars included in our analysis are located along the path
defined by the best fit. As explained in \Sec{Dtrans}, G77-61 displays larger errors
on its abundance determination. Yet, it is also located within the
envelope (square in \Fig{Dtransenv}). 
HE~0557-4840  is located at a somehow low value for [Fe/H]. According to
the exact value for \Dtrans, it could either seat outside the envelope
(\Dtrans$\simeq$-2.2) or be very close to the edge (\Dtrans$\simeq$-3).
%
%As can be seen, the normal mode of
%star formation is able to reproduce the abundances observed in the ESO-LP stars. 
%The good agreement between the observed abundances and models of global chemical evolution is
%a result of Woosley and Weaver yields \citep{1995ApJS..101..181W}, as also pointed out in \citet{2008arXiv0803.3161H}.
%
%We remind here that the predicted curve goes from high redshift (lowest
%iron enrichment) to low redshift (highest one). 

% It has to be kept in mind that our model traces the chemical state of the structures
% at each $z$, so that for models compatible with the \Dtrans\ constraints, we are able
% to provide the chemical abundances of Fe, C, and O as a function of the redshift.
% For any model, and for the best fit one in particular (used to illustrate the discussion
% in previous sections), this set a redshift of formation for the CEMP stars.

% \begin{figure}
% \includegraphics[width=\linewidth]{Dtrans_envelope.eps}
% \caption{Same as in \Fig{Dtrans}. The envelope (thin black lines)
% contains all  evolution curves predicted by models, that reproduce
% the abundance at a 2$\sigma$ level, as determined by our $\chi^2$
%  analysis
% (see text for details and references). }
% \label{f:Dtransenv} 
% \end{figure}

% \begin{figure}
% \includegraphics[width=\linewidth]{SFR_envelope.eps}
% \caption{Same as \Fig{SFRenv}. The envelope (thin black lines)
% contains all  models, that reproduce
% the \Dtrans\ evolution at a 2$\sigma$ level (\Fig{Dtransenv}).}
% \label{f:SFRenv} 
% \end{figure}

\subsubsection{Consequence for SFR and optical depth}

The envelope of the PopIII SFR for all allowed models is shown in
\Fig{SFRenv} (shaded area). Some common features can be 
noted: the massive mode becomes sub-dominant for redshift lower than 
about 15; on the contrary, it is required at a level of 2$\times 10^{-4}$ -
$3\times10^{-3}$ M$_\odot$ yr$^{-1}$ Mpc$^{-3}$ around $z\sim 24-28$. The
PopIII SFR cannot be larger than $3\times10^{-3}$ M$_\odot$ yr$^{-1}$
Mpc$^{-3}$  independently of the redshift.

%Finally, 
The envelope of the optical depth history among all models 
is shown in \Fig{tauenv} (light shaded area). 
We verify that all models are compatible with the 1$\sigma$ limit
of  WMAP5.

% \begin{figure}
% \includegraphics[width=\linewidth]{tau_envelope.eps}
% \caption{Same as bottom panel of \Fig{tau}. The envelope (thin black
%  lines) shows the maximum and minium values of optical depth reached by
%  the different models that reproduce
% the abundance at a 2$\sigma$ level (see also \Fig{SFRenv} and \Fig{Dtransenv}).}
% \label{f:tauenv} 
% \end{figure}

\begin{figure}
\includegraphics[width=\linewidth]{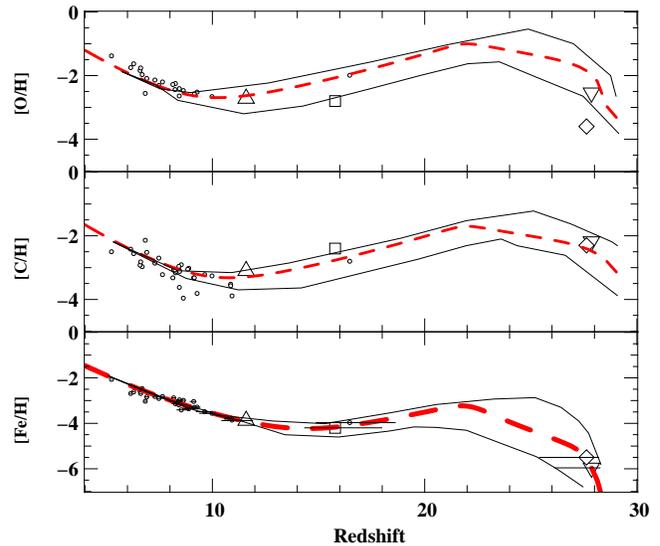}
\caption{Evolution of the [Fe/H], [C/H] and [O/H] abundances with redshift. Dashed
red curves and black solid lines correspond to the prediction from the
 best model and to the 95\%~CL envelopes (\Sec{grid}). The SFR of the
 best model is shown in \Fig{SFR}. Symbols used for the stars are the
 same as in \Fig{Dtrans}. Horizontal lines in the bottom panel give the
 estimated errors on redshift for CEMP stars (\Tab{zout}).}
\label{f:abundances} 
\end{figure}

\subsubsection{Redshift of formation}

%Finally, 
As explained in \Sec{Ztoz}, for a given model, we have assigned a typical redshift, $z_\star$, of
formation to each individual star. % in order to best fit \Dtrans\ and [Fe/H] simultaneously. 
Thus, we can locate each star on a plot of abundances versus redshift, and compare them directly to the predicted evolution. 
This is done in \Fig{abundances} using the best fit model (dashed red
line).  
We also show the corresponding 95\%~CL envelopes
for the evolution of the abundances (thin black lines). Those envelopes 
also bracket the errors on the redshift of formation (see [Fe/H] plot 
in \Fig{abundances}). The redshift ranges are given in \Tab{zout}.

We see that CEMP stars are likely to form at high redshift.
For example, the star HE~1300+0157 (upper triangle in \Fig{Dtrans}) is found
%, from the minimization of
%$\chi^2(\star,z,p)$ 
to be formed at $z_\star\sim 11$ (for the best fit
model). HE~1300+0157
is located along a branch of the evolution where the dilution by IGM
accretion is the only important process.  Then, abundances decrease
very slowly and uncertainties on the redshift are  of the order of 100 Myr
($15<z<20$). It is interesting to note that the predicted abundances 
also reproduce the Fe, C and O abundances at this specific redshift. 
The star G 77-61 (square), that has a similar iron abundance but a higher carbon
abundance,  is born earlier
 (it is located on a different place along 
the curve \Dtrans-[Fe/H]; \Fig{Dtrans}) at $z\sim 16$.
The exact redshift for HE~1327-2326 and HE~0107-5240 (lower
triangle and diamond) depends on the exact time
of formation of the first structure, which is largely unknown. The
duration of this episode is only a few tens of Myr, which also
corresponds to  the
uncertainty on the redshift of formation of those stars. 
Finally, all stars observed in the ESO-LP are found to form at
redshift below 10, except CS 22949-037 whose abundances are very similar
to HE~1300+0157 (\Sec{Dtrans}).

\subsection{Uncertainties}

\subsubsection{Peculiar stars}

The observed oxygen abundance in the UMP star HE~0107-5240
(diamond in upper panel of \Fig{abundances}) seems to be lower than the predicted abundances.
This discrepancy   points to the very special stars where C, O
abundances are not explained by standard predictions \citep{2006NuPhA.777..424N,2007ApJ...660..516T}.
%, or to a likely limitation of our approach which only deals with
%average quantities.
It is clear that the yields of 
the massive stars at zero metallicity are
poorly known. For example, new calculations concerning the
 effect of rotation on the
evolution of primordial stars can modify considerably the abundance of C
and O, and their corresponding ratio \citep{2008A&A...489..685E}.
They would then allow for a decrease of about 1 dex required for the oxygen
abundance of HE~0107-5240 to be reproduced.

\subsubsection{1D vs 3D model atmospheres}
\label{s:1d3d}

We now study the influence of uncertainties on stellar abundances on the overall results.
We have performed the same full analysis using the abundances derived
from 1D model atmospheres (given in
\Tab{CEMP}), yielding  new envelopes for the \Dtrans-[Fe/H] plane,
abundance evolution, optical depth evolution and SFR of PopIII stars.
In the following figures, the same symbols as above are used for each CEMP
star. We do not show ESO-LP stars in the figures, as their positions are
unchanged.

\begin{figure}
\includegraphics[width=\linewidth]{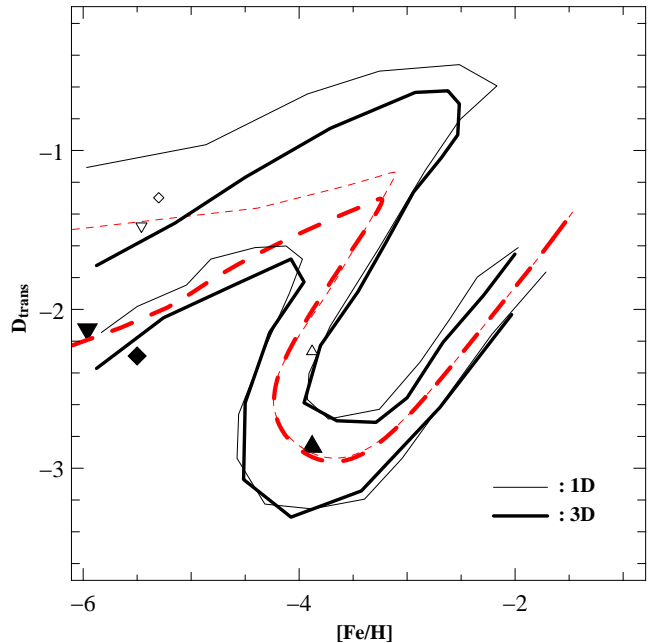}
\caption{The transition discriminant, $\Dtrans$, as a function of
[Fe/H] (same as \Fig{Dtrans}). The abundances of CEMP stars
derived from 1D and 3D model atmospheres (\Tab{CEMP}) are marked with filled and open symbols respectively.
 The 95\%~CL envelopes  of predicted evolution assuming 3D
 and 1D model atmospheres  are displayed with thick and thin solid lines
 respectively, while best models are shown with thick and thin dashed lines.}
\label{f:Dtrans1d3d} 
\end{figure}

The main effect of the correction applied to  abundances  derived
from 1D  model atmospheres is to lower all abundances by about
0.1 dex for [Fe/H], and up to 1 dex for C and O abundances. This can be
seen  in \Fig{Dtrans1d3d} by comparing filled and open symbols
(abundances derived from 3D and 1D model atmospheres respectively). Given the predicted path of \Dtrans-[Fe/H],
this does not modify much the envelope in the intermediate redshift
 range. At high redshift, the effect is strong and displaces the envelope
towards lower values of \Dtrans. 
In both cases, PopIII stars are required and a very similar  pattern
is displayed. 

As for the SFR, the best fit model assuming 
abundances derived from 1D model atmosphere extends slowly down to about $10^{-7}$ M$_\odot$ yr$^{-1}$
Mpc$^{-3}$ at $z\sim4$.
 However, the global envelope (including all possible shapes)
is almost identical in both cases, although a bit more extended assuming
1D model atmospheres. The evolution of optical depth is also not modified in a significant way, since the contribution of PopIII stars to reionization is marginal.

\begin{figure}
\includegraphics[width=\linewidth]{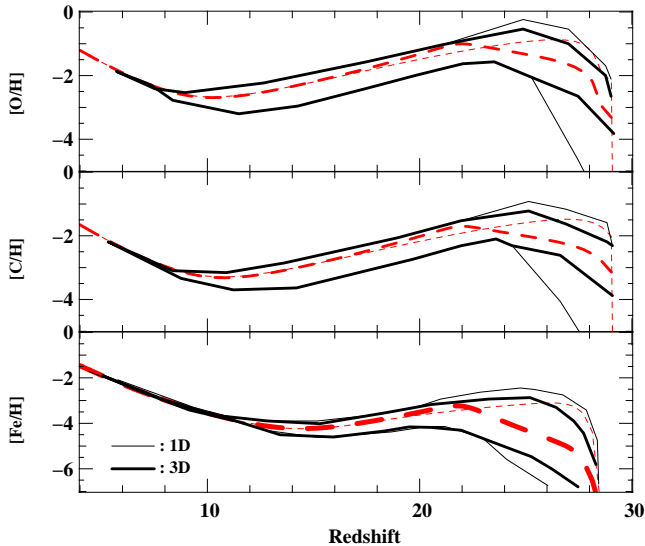}
\caption{Evolution of the [Fe/H], [C/H] and [O/H]  abundances
 with redshift (same as in \Fig{abundances}). The 95\%~CL envelopes of allowed
 evolutions assuming abundances derived from 3D and 1D model atmospheres
 for CEMP stars are displayed with thick and thin lines
 respectively. The envelope as well as the 
evolution for the best models (thick and thin dashed red lines) are quite different
at high redshift only. }
\label{f:abundance1d3d} 
\end{figure}

Finally, we consider how our estimates for the redshift of
formation of CEMP stars are affected. 
\Fig{abundance1d3d} displays the envelope for the  evolution of
abundances with redshift  allowed in the two cases. It appears even more
clearly that the discrepancy occurs only at high redshift, while the
transition from CEMP to standard metal-poor stars is required in both cases.
The stars are not located in this figure for clarity. 
 The different ranges of estimates for their redshift of formation are summarised
in \Tab{zout}.  
Again, only the two stars at the highest redshift are affected by this
uncertainty;  while the range of allowed redshift is larger in the case of abundances
derived from 1D model atmospheres (this is not visible in \Fig{abundance1d3d}).

\begin{table}
\caption{{\bf Estimation of the redshift of formation of CEMP stars}}
\label{t:zout}
\begin{tabular}{ccccccc}
Name & \multicolumn{3}{c}{3D model$^b$} & \multicolumn{3}{c}{1D model$^b$} \\
     &    $z_{\rm min}^a$& $z_{\rm best}$& $z_{\rm max}^a$ & $z_{\rm min}^a$&    $z_{\rm best}$&    $z_{\rm max}^a$ \\
\hline
HE~1300+0157   & 10.4  & 11.6  & 11.9 &9.7   & 11.5  & 12.0 \\
HE~1327-2326   & 26.1  & 27.9  & 28.3 &24.9  & 28.3  & 28.4 \\
HE~0107-5240   & 25.4  & 27.6  & 28.2 &24.7  & 28.3  & 28.3 \\
G-77-61        & 14.3  & 15.8  & 18.0 &14.3  & 15.9  & 19.5 \\
\hline
\end{tabular}
$^a$ Minimum and maximum redshift predicted by models that match
 abundances of all stars at a 95\%~CL (\Sec{grid}). $^b$ Abundances are
 derived from 3D model atmospheres (second to fourth columns) or from 1D
 model atmospheres (fifth to last columns).
\end{table}

%%%%%%%%%%%%%%%%%%%%%%%%%%%%%%%%%%%%%%%%
%%%%%%%%%%%%%%%%%%%%%%%%%%%%%%%%%%%%%%%%
\section{Discussion and conclusions}
\label{s:conclusion}

%model used
Following the approach developed in \citet{2004ApJ...617..693D}, we have
modeled  the evolution of individual element abundances in the ISM assuming homogeneous star formation and stellar
yields. Recent observations at z$\sim$ 7-8 \citep{2007ApJ...670..928B} were used to better constrain
one ingredient of the model, namely the SFR for PopII/I at high redshift. 
We have shown that a
homogeneous scenario of hierarchical structure formation reproduces
many different  observations, from reionisation and first star
abundances to local abundance observations.
% optical depth
We found that using
the most recent results on the optical depth from WMAP, a massive mode
is not absolutely required. Nevertheless, the data can accommodate a PopIII contribution, responsible for the
gradual reionization starting from $z\simeq 20$.  Although the cosmological importance  of PopIII stars
cannot be fully constrained by the integrated Thomson optical depth, this question may be
better tackled in the future with the help of accurate measurements of the CMB polarization
data \citep{2009ApJS..180..306D}.

% stellar constraints
We have also considered stellar constraints, in particular the MDF and the evolution of
\Dtrans\ with [Fe/H]. 
% MDF
In the literature, the slope of the MDF can be reproduced by different models of chemical
enrichment, from galactic models to hierarchical analytic models including merger trees, as
shown in Fig. 12 of \citet{2008arXiv0809.1172S} (see also
\citealt{2006ApJ...641....1T}).  
This indicates that the slope of the  MDF does not discriminate between the methods used and the
associated   level of heterogeneity. In contrast, the pattern at [Fe/H]$<-3.5$ is more
difficult to reproduce. \citet{2008arXiv0809.1172S} claim that no model considered in
their paper can reproduce the MDF at very low iron abundances. We find that the modification
of the tail at low values of [Fe/H] may be  related to the presence of  PopIII stars.  In our study, we have tested the impact  of the massive mode by varying
the typical PopIII minimal mass from 8 to 40 \msun. This appears to be an important parameter as
found in the study of \citet{2007MNRAS.381..647S}. They used a mass range from 140 to
200~\msun\ which we consider disfavoured by the very specific yields of the stars within this mass range \citep{2006ApJ...647..773D}. In this paper, we demonstrate that the minimum mass must be larger than about 30
\msun\ in order to reproduce the observed part of the MDF. 

%CEMP
We have shown that the existence of  PopIII stars at high redshift is required to explain the
abundance pattern observed in the CEMP stars. In addition, we have shown that a massive mode with
a typical mass of  40 \msun\ reproduces the evolution of observed \Dtrans. In contrast, it
is known that PISN with masses 140-200~\msun\ do not provide the correct chemical pattern
and cannot reach high values of \Dtrans\ for low values of [Fe/H] (the ratio C/Fe in their
yields, \citealt{2003ApJ...591..288H}, is not high enough). However, if  it could be established that the
CEMP stars were particular cases, such as belonging to binaries \citep{2005ApJ...635..349R,2007ApJ...665.1361T} or due to the
preferential depletion of iron in grains \citep{2008ApJ...677..572V},
the SFR related to PopIII
stars would be diminished, at least as far as chemical evolution is concerned. 

In conclusion, our analysis  hints at a massive mode at
$z\simeq 20-30$, which becomes sub-dominant at lower $z$ ($z\sim 15$).\\

Going further requires improvements from the observational and theoretical
point of view. 
On the theoretical part, the existence of PopIII stars with masses of a few tens of solar masses is also suggested by
recent hydrodynamical simulations 
\citep{2003Natur.425..812B,2006MNRAS.373..128G,2006MNRAS.366..247J,2007MNRAS.374.1557J,2007ApJ...665...85J,2007ApJ...663..687Y,2008MNRAS.387.1021G,2009ApJ...691..441S}. The explosion of these high mass stars disrupts the surrounding environment and delays the
formation of lower mass stars  {\em within the same structure}.  Merger trees
\citep{2007MNRAS.381..647S} account for such a mechanical feedback which delays  PopII/I star
formation along a single branch.  In a homogeneous picture, this delay translates into a
reduction of the SFR between the two modes. We note that  this epoch is very short in our
model (100-200 Myr). However,
the exact evolution of the structures are still
uncertain.
One also needs  improved yields provided by stellar
evolution models at all metallicities  and, more specifically, 
 SN calculations that are even more uncertain.
 Note however that uncertainties related to the
determination of abundances (derived from 1D or 3D model atmospheres) do not change our
conclusion on the cosmological importance of PopIII stars.

Many improvements are possible in the future with regard to  observations.
 $(i)$ Additional stellar constraints and abundance measurements. Better statistics are
necessary to construct the MDF at very low metallicity, with next
 generation optical telescopes such as GMT. It will then be
possible to constrain in a better way 
the epoch of  the massive population (whose duration is related  the critical
metallicity as used in merger trees;  \citealt{2007MNRAS.381..647S,2006ApJ...641....1T})  and  the typical
mass range of PopIII stars. $(ii)$  More complete spectroscopic observations in very
metal-poor stars are needed in order to obtain a complete set of observations in the 
\Dtrans\ diagram. Of particular importance  is the continued search for
 HMP stars and UMP stars to test
their statistical significance relative to the bulk of the VMP/EMP stars. 
$(iii)$ In future studies, one could attempt to better constrain the model using  other chemical
elements, in particular r and s-process elements which can bring new
 constraints on the mass range of massive stars, PopIII \citep[e.g.][]{2008arXiv0812.1227F}.
$(iv)$ Additional cosmological constraints at high
redshift. Massive objects may be directly observable, as more and more quasars and galaxies
are detected at high redshift
\citep{2003AJ....125.1649F,2004ApJ...607..697K,2006Natur.443..186I} and
with JWST in the future (e.g. \citealt{2009AAS...21342603S}).\ However, the extreme brevity
of the PopIII epoch makes the connection unclear \citep{2007AAS...211.9114A}. A more promising probe
would be the observations of high redshift gamma-ray bursts, that correspond, for the longer
bursts, to the deaths of massive single stars
\citep{2003ApJ...591..288H,2004ApJ...604L...1B,2006ApJ...642..382B,2006MNRAS.372.1034D}. Recent
 observations of GRBs at $z>6$ (GRB 050904 at $z=6.3$, \citealt{2006Natur.440..184K}; GRB 080913 at $z=6.7$, \citealt{2009ApJ...693.1610G}) 
and even $z>8$ (GRB 090423 at $z\sim 8.2$, \citealt{2009arXiv0906.1577T})
are  extremely encouraging.
$(v)$  In addition, massive stars would have
polluted their environment with an initial enrichment of heavy elements which could be
compared to the one observed  in the Ly$\alpha$ forest along quasar absorption spectra at
$z\lsim 6$ \citep{2001ApJ...561L.153S,2002ApJ...576....1A,2004A&A...419..811A} or in the Damped Lyman $\alpha$ systems
\citep{2003MNRAS.346..209L,2003ApJ...595L...9P}. This requires a better understanding of the outflows of
metals into  the IGM which is beyond the scope of this work. In addition, it will be
important to better understand the role of inhomogeneities  on the different populations of
the first stars. The uniformity and extent of the metal pollution is also under debate, and
could in the future be used to distinguish between local and recent pollution and global
pollution by an earlier population of stars \citep{2001ApJ...555...92M,2002ApJ...571...40M,2003ApJ...591...38W}.

\section*{Acknowledgments}

We are very grateful to Roger Cayrel and Patrick Petitjean for their always pertinent and
fruitful comments. We also thank Sylvia Ekstr\"om for useful discussions
on the yields of PopIII stars.
We thank  the referee very much  for a very careful reading and advice specifically regarding observational aspects.
We thank E. Thi\'ebaut, and D. Munro for freely  distributing his Yorick programming language
({\em \tt http://yorick.sourceforge.net/}), which we used to implement our analysis.  
The work of E.V. and K.A.O. has been supported by the collaboration INSU-CNRS France/US.
The work of K.A.O. was partially supported by DOE grant DE-FG02-94ER-40823. 

\label{lastpage}

\bibliography{RVMODSV_MNRAS_3}
\end{document}